# POPDx: An Automated Framework for Patient Phenotyping across 392,246 Individuals in the UK Biobank Study

Lu Yang[1*], Sheng Wang[4] and Russ B. Altman[1,2,3*]

[1]Department of Bioengineering, Stanford University, Stanford, CA 94305, USA

[2]Department of Genetics, Stanford University, Stanford, CA 94305, USA

[3]Department of Medicine, Stanford University, Stanford, CA 94305, USA

[4]Paul G. Allen School of Computer Science and Engineering, University of Washington, Seattle, WA 98195, USA

*To whom correspondence should be addressed.

**ABSTRACT**

**Objective**

For the UK Biobank standardized phenotype codes are associated with patients who have been hospitalized but are missing for many patients who have been treated exclusively in an outpatient setting. We describe a method for phenotype recognition that imputes phenotype codes for all UK Biobank participants.

**Materials and Methods**

POPDx (Population-based Objective Phenotyping by Deep Extrapolation) is a bilinear machine learning framework for simultaneously estimating the probabilities of 1,538 phenotype codes. We extracted phenotypic and health-related information of 392,246 individuals from the UK Biobank for POPDx development and evaluation. A total of 12,803 ICD-10 diagnosis codes of the patients were converted to 1,538 phecodes as gold standard labels. The POPDx framework was evaluated and compared to other available methods on automated multi-phenotype recognition.

**Results**

POPDx can predict phenotypes that are rare or even unobserved in training. We demonstrate substantial improvement of automated multi-phenotype recognition across 22 disease categories, and its application in identifying key epidemiological features associated with each phenotype.

**Conclusions**

POPDx helps provide well-defined cohorts for downstream studies. It is a general purpose method that can be applied to other biobanks with diverse but incomplete data.

## BACKGROUND AND SIGNIFICANCE

Artificial intelligence (AI) allows machines to recognize patterns in electronic patient records (medical notes, laboratory tests, medications, and diagnosis codes). With increasing amounts of data available, machine learning algorithms have enabled healthcare applications, ranging from the detection of pneumonia in frontal chest X-ray images to the identification of heart failures in clinical notes [1,2]. There have also been growing efforts to predict clinical events, i.e., the automatic prediction of patient phenotypes with data-driven approaches [3,4]. However, most studies have focused on a small number (<10) of disease diagnoses (e.g., assessing the risks for cardiovascular diseases), and so their general utility is limited [5]. Large-scale biobanks with genetic and phenotypic data are a vital source for studying a wide range of diseases. Cohort studies such as UK Biobank support broad multi-phenotype research with a range of data including biological samples, physical measures, questionnaires related to socio-demographic conditions, lifestyle and health-related factors, and electronic medical records [6–8]. Unfortunately, missing data is common. In the UK Biobank, many individuals who have been treated exclusively on an outpatient basis have missing phenotype labels. To maximize the utility of these data, large-scale patient phenotyping is necessary but expensive, time-consuming, and difficult. Currently, only a subset of conditions have available algorithms for recognition of unlabeled phenotypes [9–11]. These algorithms require extensive task-defined preprocessing and ad-hoc feature engineering [7,8,11]. A disease recognition system that recognizes multiple phenotypes would be helpful in defining patient cohorts for downstream studies. Recognizing rare phenotypes with small (or non-existent) training data is a particular challenge, even in large biobanks.

Rare diseases affect about 3.5–5.9% of people worldwide [12]. While predictive models exist for common diseases using carefully curated datasets in sufficient volume to allow statistical characterization, detecting rare or unseen diseases remains difficult [13,14]. There is currently no framework that evaluates individual patients for rare and common diseases in parallel. Rare diseases can be associated with noisier data because of inconsistent diagnostic criteria and clinician uncertainty [15]. The phenotype-driven approaches to rare diseases therefore typically rely on difficult-to-assemble cohorts. For rare diseases, patient sample sizes follow a long-tailed class distribution. Conventional machine-learning methods typically perform better on the majority class and exhibit poor predictive accuracy on rare disease classes. In recent years, semi-supervised and supervised methods have helped improve performance on imbalanced datasets, e.g., single-cell annotations to classify cells into cell types and cell states absent or absent in the training data [16,17]. However, the techniques they employ have not been applied to multi-phenotype recognition with heterogeneous patient data. We developed POPDx to associate patients with phenotypes for both common and rare phenotypes. It combines embedded representations of disease features with NLP-based encoding of the text and network-based embedding of the Human Disease Ontology to regularize the disease feature representation. We train POPDx with numerical and categorical data including health records, laboratory tests, individual demographics, lifestyles, and environmental exposures. We compile clinical profiles of 392,246 patients in UK Biobank [6] and perform imbalanced learning with 1,538 disease and health-related labels. Our phenotype recognition algorithm outperforms the state-of-the-art predictive models. It recognizes a comprehensive set of phenotypes, and makes the following contributions:

1. It manages missingness, noise, and high dimensionality typical in the electronic health record (EHR) data.

2. It scales to population-scale sets of patients and phenotypes.

3. It leverages the Human Disease Ontology to derive an integrated model for 1,538 phenotypes, that can recognize phenotypes even when there are few or no examples of these phenotypes in the training set.

## MATERIALS AND METHODS

The POPDx framework leverages phecode embeddings that are constructed from the Human Disease Ontology covering all the diagnostic codes in UK Biobank and the textual descriptions of the phenotypes [18–21] to achieve simultaneous recognition of multi-phenotype that outperform the state-of-the-art models. We assessed our embeddings by computing the dissimilarity of phenotypes within and outside of the disease category. The importance scores of 38,663 features were evaluated to aid the POPDx explainability.

### UK Biobank cohort

We extracted phenotypic and health-related information from the UK Biobank including clinical assessments, lifestyle questionnaires, physical measurements, and electronic medical records. Among approximately 500,000 individuals from the UK Biobank dataset, 392,246 individuals have ICD-10 coded diagnosis information. We binned and applied one-hot encoding to numerical and categorical features, respectively, into 38,663 binary variables. For 1,538 diagnostic labels, we map 12,803 International Classification of Diseases Tenth Revision (ICD-10) codes to 1,538 phecodes [18–21]. ICD billing codes are routinely used to identify patient cohorts from large observational datasets. Cases of multiple ICD codes are often accumulated to define the case or control status of a specific phenotype [21]. The use of phecodes is one recognized strategy in clinical research that combines relevant ICD codes into meaningful phenotypes [22]. For repeatability of POPDx to be established, we leveraged a beta version of map from ICD-10 to phecode introduced by Wu et al. which was validated based on about 84% coverage of the ICD-10 codes in the UK Biobank database [18]. The entire dataset was split into training, validation, and test sets. Instead of dividing the data randomly, we split the individuals to allow experimental evaluation of unseen, rare, and common diseases. We generated a large multi-label patient dataset to contain phecodes that are present or absent in the training dataset. Some phecodes occur zero or a few times to simulate unseen and rare diseases while others are common.

**The joint semantic and structure-based embeddings of phenotypes**

We relied on both the textual information and hierarchical tree-structure of the phenotypes to compute the joint semantic and structure-based embeddings of phecodes. The phecode embeddings depend on embeddings of the ICD-10 codes which comprise them. We downloaded the hierarchical tree-structured representation of the ICD-10 data from the online showcase of UK Biobank resources. We use the hierarchical relation of the ICD-10 codes to construct an undirected network of phenotypes. There are 19,155 nodes in the network, corresponding to 19,155 codes. Edges are not weighted. We perform a shortest-path graph search and then compute the low-dimensional representation of each diagnostic code by using the singular value decomposition (SVD) [23]. We thus have a compressed, low-dimensional representation of each ICD-10 code based on the undirected disease network.

The text description of each ICD-10 code is a sentence or a short term that characterizes the meaning of the code. For example, ICD-10 code P29 is described as "*Cardiovascular disorders originating in the perinatal period*". We use the pre-trained version of the BioBERT [24] model which has been widely adopted and very effective for biomedical text mining tasks, to extract a fixed vector of ICD-10 code based on this text. The BERT [24,25] model breaks down the textual description of each ICD-10 code into tokens. Then, it adds a special classification token [CLS] at the beginning of each text. The 768-dimensional hidden state embedding of the "[CLS]" token from the last layer is used as the aggregate representation for the ICD-10 code. Finally, we merge the final two representations from NLP-mapping and network-based embeddings of disease terms. Given the ICD-10 codes and the vector representation of their textual description, we can calculate the embedding for each phecode by averaging the representations of the ICD-10 codes to which the phecode corresponds. We included 12,803 ICD-10 codes present in our UK Biobank dataset which map to 1,538 phecode labels (Figure 1B, C).

**Calculating phenotype dissimilarity**

We compute three measures of phecode dissimilarity based on the disease ontology embedding, the embedding of the associated text, and both together. We measure the distance between embeddings with cosine distance. The distances are calculated as in-group (intra-) and out-group (inter-) distances. The cosine distances between a phenotype and those within the same disease category are considered in-group dissimilarity. The cosine distances between phenotypes of different disease categories are computed as out-group dissimilarity.

**Simultaneous recognition of multiple-phenotypes**

Our algorithm (Figure 1B) leverages the text description and ontological relationships of phenotypes to predict novel phenotypes (with no training examples) by relating them to clinically and contextually similar phenotypes. We use a bilinear model to predict the disease type for both seen and unseen phenotypes. Let $P$ be an $m$ by $n$ matrix of input embedding of the patients, where $m$ is the number of patients and $n$ is the number of features. Let $Y$ be an $m$ by $c$ label matrix, where $c$ is the total number of phenotypes. $Y_{ij}$=1 if patient $i$ has a diagnostic label of phenotype $j$, otherwise $Y_{ij}$=0. c is the total number of phecodes, and the majority of these phenotypes have fewer than 1000 examples in the training data. For example, when a patient is associated with 126 disease labels, the corresponding columns of diseases are ones while the others are all zeros in the label matrix. Let $U$ be a $c$ by $h$ matrix of the low-dimensional representations of disease types, where $h$ is 1,268, the dimension of phenotype embedding space. We optimize the following binary cross-entropy loss:

$$\sum_{i=1}^{m} \sum_{j=1}^{c} \left[ Y_{ij} \log \sigma\left( P_i W_1 W_2 U_j^T \right) + \left( 1 - Y_{ij} \right) \log \left( 1 - \sigma\left( P_i W_1 W_2 U_j^T \right) \right) \right]$$

where $W_1 \in R^{n \times q}$ and $W_2 \in R^{q \times h}$ are the parameters that need to be estimated, and $q$ is set to be 150 through parameter tuning. POPDx optimizes the objective function by Adam optimizer. After the optimization, the likelihood of a diagnostic code $j$ presented by a patient with a feature vector $p$ is estimated as

$$L_j = Sigmoid\left( p W_1 W_2 U_j^T \right)$$

where $L_j$ is the probability that the phenotype $j$ belongs to this patient. $L$= {$L_1$, $L_2$, ..., $L_{1538}$} is the probability distribution of diagnosis labels for a patient. We use Pytorch [26], Matplotlib [27], and Numpy [28] for the experiments.

**Feature importance analysis**

Scoring the importance of individual features provides some interpretability for model predictions. We use DeepLIFT to compute the importance and relevance of 38,663 features on each of the 1,538 phenotypes via a balanced selection of true-positive and true-negative cases. DeepLift is a backpropagation algorithm that measures the contribution of individual features on the output of a neural net for a specific input [29]. It computes the differences between the activation of each neuron and their reference activation, where the "reference" is computed based on the selected negative samples. DeepLIFT highlights both positive (supportive) and negative (not supportive) influences on the prediction. The magnitude of the relevance value corresponds to its importance. We implemented DeepLIFT in the framework Captum [30]. First, DeepLIFT scores are computed for each feature and for each patient. Then, for each phenotype, the DeepLIFT scores of all the true-positive patient data are averaged to obtain an importance score for each feature. We create a vector of feature importance scores for each phenotype.

**RESULTS**

**Overview of POPDx**

Figure 1A summarizes POPDx. First, the raw data are downloaded from UK Biobank. Second, the collected data are transformed into 38,663 patient features and 1,538 associated phenotype codes. Third, we apply POPDx to recognize a diverse set of phenotypes, yielding a profile of phenotypes for each patient. Because the training data for many phenotypes are sparse, we introduce the use of ontological relationships to supplement the raw data. In particular, POPDx framework leverages disease ontological relationships (as represented in the Human Disease Ontology) embedded in a low-dimensional space and then projects the high-dimensional features of each patient to the same low-dimensional embedding space by a nonlinear transformation (Figure 1B and 1C). This has been used in other settings and has been shown to improve classification for classes with zero or few examples [16]. The framework encodes the patient data through a bilinear framework with two hidden layers of POPDx architecture and a matrix transformation (Figure 1B). The resulting outputs denote the probabilities of each phenotype for each patient. POPDx is written in Python and is made available as an open-source package. Importantly, with a pre-trained model, we can recognize 1,538 disease phenotypes given an input patient matrix in a few minutes on a GPU.

**392,246 Individuals Selected from UK Biobank cohort**

In the UK Biobank, there are about 500,000 participants in total. 392,246 of these individuals have ICD-10 codes in their records. For these patients, we selected 219,604, 86,361, and 86,361 unique patients for training, validation, and testing respectively. The three sets have similar basic characteristics (Table 1). Fifty-six percent of individuals are women. The majority of participants are white. Elderly adults dominate the selected cohort with an average age of 71. We assessed 1,538 phecode labels extracted from 12,803 ICD-10 codes from the cohort. Diagnostic labels have a long-tail distribution (Figure 2A): nearly 40% of these phecode labels have fewer than 100 positive patients (Figure 2C). Among 392,246 individuals, 377,612 people have fewer than 30 phenotypes (Figure 2B). We integrate 38,663 category-specific features and summarize them into 20 data subgroups as potential risk factors to aid our analysis (Table S1).

Table 1. Basic characteristics of the selected UK Biobank cohort

| | Training set (N = 219,604) | Validation set (N = 86,321) | Test set (N = 86,321) |
|---|---|---|---|
| **Race, n%** | | | |
| White | 206,394 (93.985) | 81,646 (94.584) | 81,365 (94.259) |
| Mixed | 1,304 (0.594) | 478 (0.553) | 497 (0.576) |
| Asian | 3,489 (1.589) | 1,394 (1.615) | 1,364 (1.580) |
| Black | 4,459 (2.030) | 1,468 (1.701) | 1,621 (1.878) |
| Chinese | 728 (0.332) | 159 (0.184) | 213 (0.247) |
| Other | 2,046 (0.932) | 670 (0.776) | 747 (0.865) |
| Unknown | 690 (0.314) | 293 (0.339) | 325 (0.377) |
| Did not answer | 494 (0.225) | 213 (0.246) | 189 (0.219) |
| **Sex, n%** | | | |
| Female | 122,353 (55.715) | 47,315 (54.813) | 47,786 (55.358) |
| Male | 97,251 (44.285) | 39,006 (45.187) | 38,535 (44.641) |
| **Age, n%** | | | |
| < 65 years | 58,988 (59.447) | 16,905 (59.472) | 19,680 (59.449) |
| ≥ 65 years | 160,616 (74.378) | 69,416 (75.381) | 66,641 (74.946) |
| All | 219,604 (70.367) | 86,321 (72.266) | 86,321 (71.413) |

**Phecode embeddings reflect disease similarity**

Since POPDx addresses rare and unobserved diagnostic codes based on their textual description and the ontological relationship to common diseases, its performance relies on high quality embeddings of phenotypes. For that reason, we verified that diagnostic codes that are direct neighbors in the graph of Human Disease Ontology are also close in our low-dimensional embedding space. We compared three types of phenotype similarities: the disease ontology structure-based similarity, the text-based similarity, and joint semantic and structure-based similarity (**Methods**). We assessed the phenotype embeddings using direct neighborhood and non-direct neighborhood proximity. We first observe the average cosine distance of direct neighbors in the disease ontology graph is $0.15 \pm 0.04$, which is 69.70 % higher than that of the text-based embeddings ($0.05 \pm 0.01$), while the average cosine distance of k-hop neighbors ($0.34 \pm 0.02$) in the disease ontology graph is 83.02% higher than those of the text-based embeddings ($0.06 \pm 0.01$) (Figure 3B). The average cosine distances of joint embeddings in the same disease type neighborhood and k-hop neighborhood are $0.05 \pm 0.01$ and $0.07 \pm 0.01$ respectively (Figure 3C). POPDx incorporates the joint semantic and topology preserving embeddings under the principle that the unobserved or unseen phenotypes in training can borrow information from other disorders based on their shared characteristics.

Dimension reduction via $t$-distributed stochastic neighbor embedding ($t$-SNE) [31] on the joint semantic and structure-based embeddings reveals distinct disease groups (Figure 3A). The joint embeddings of phecodes via a biomedical domain-specific pre-trained language model [24] and canonical classification of the diagnoses present disjoint clusters in the low-dimensional embedding space (Figure 3A, Figure S4 A, B). Whereas diseases of most organ systems (Figure 3A, B) are independently clustered in latent space (for example, Diseases of the ear and mastoid process, Diseases of the eye and adnexa, Diseases of the circulatory system), some categories of diagnostic codes do not form a clear cluster but intermix (e.g., External causes of morbidity and mortality, Injury, poisoning and certain other consequences of external causes). To quantify the similarity of the phecodes within and outside of the disease category, we measure the two types of dissimilarity of each disease class (Figure S2). The phenotype codes for pregnancy, childbirth, and the puerperium have the highest similarity with a mean in-group cosine distance of $0.04 \pm 0.01$, in contrast with diseases of the musculoskeletal system and connective tissue that have the lowest in-group similarity with a mean in-group cosine distance of ($0.08 \pm 0.06$). The relationships are visualized in the dendrograms (Figure S3) which present the hierarchical relationship between our phenotypic embeddings for different disease categories. For example, phenotypes for the infectious and parasitic diseases (Figure S3A) form correlated groupings in the topological space, such as 41.1 and 41.2 (staphylococcus and streptococcus infections).

**Improved disease recognition for unseen, rare, and common phenotypes**

To evaluate the performance of POPDx on different phenotypes, we categorize phenotypes according to the number of instances in the training set (Figure 4). Most diagnostic labels (98.6%) have fewer than 10,000 patient samples (Figure 2C). POPDx (Figure 4, S6) yields AUROC (Area Under the Receiver Operating Characteristic Curve) scores of 0.71 and 0.74 for phenotypes with 0-10 and 10-100 training samples. The AUROC and AUPRC (Area Under the Precision-Recall Curve) scores improve with training size. We also investigated performance for phenotypes with no patient sample in the training dataset. For these, POPDx achieves an AUROC score of 0.68 and an AUPRC score of 0.24, which are 74% and 218% higher than those of the logistic regression baseline. A sampling ratio of positive to negative patients of 1:10 is consistently used to report the AUROC and AUPRC of all the experiments.

Across different disease categories of phenotypes, we investigate how well POPDx outperforms the baseline of logistic regression based on AUROC and AUPRC scores (Figure 5A, B). We achieve an AUROC of 0.81 and an AUPRC of 0.37 with 131 phenotypes for Diseases of the circulatory system (Table 2). We outperform the random forest and logistic regression for increases in AUROC and AUPRC scores by 0.16 and 0.15. We compared POPDx with other strategies for phenotype embedding (Figure 4, Figure 4S). The joint semantic and structure-based embedding method achieves the best performance compared to NLP-based and ontology-based frameworks. Interestingly, for phenotypes that have fewer than 100 patient samples, the NLP-based and ontology-based frameworks improve the AUROC and AUPRC scores compared to those of random forest and logistic regression. To assess the ability of POPDx to work with even larger sets of phenotypes, we applied patient phenotyping across 12,803 ICD-10 diagnostic codes (almost 8 times more codes than with phecodes) with the same dataset. POPDx detects diagnostic labels for both rare and common codes with competitive AUROC scores (Figure 5S).

Table 2. AUROC and AUPRC scores of different disease categories

| Disease Category (Abbrev.) | AUROC | | AUPRC | |
|---|---|---|---|---|
| | Mean | SD | Mean | SD |
| (PERIN) Certain conditions originating in the perinatal period | 0.6859 | 0.1669 | 0.2606 | 0.2253 |
| (ID) Certain infectious and parasitic diseases | 0.7215 | 0.1106 | 0.2793 | 0.1318 |
| (CONGEN) Congenital malformations, deformations and chromosomal abnormalities | 0.6403 | 0.1686 | 0.1953 | 0.1156 |
| (BLOOD) Diseases of the blood and blood-forming organs and certain disorders involving the immune mechanism | 0.7520 | 0.0649 | 0.2851 | 0.1105 |
| (CV) Diseases of the circulatory system | 0.8141 | 0.0725 | 0.3695 | 0.1238 |
| (GI) Diseases of the digestive system | 0.7913 | 0.0887 | 0.3621 | 0.1495 |
| (EAR) Diseases of the ear and mastoid process | 0.7561 | 0.1007 | 0.3179 | 0.1598 |
| (EYE) Diseases of the eye and adnexa | 0.7642 | 0.0762 | 0.3132 | 0.1103 |
| (GU) Diseases of the genitourinary system | 0.7902 | 0.0765 | 0.3252 | 0.1177 |
| (MSK) Diseases of the musculoskeletal system and connective tissue | 0.7423 | 0.0982 | 0.2819 | 0.1285 |
| (NEURO) Diseases of the nervous system | 0.7560 | 0.0849 | 0.3232 | 0.1592 |
| (RESP) Diseases of the respiratory system | 0.8113 | 0.0700 | 0.3886 | 0.1434 |
| (SKIN) Diseases of the skin and subcutaneous tissue | 0.6995 | 0.1149 | 0.2439 | 0.1392 |
| (ENDO) Endocrine, nutritional and metabolic diseases | 0.7633 | 0.1135 | 0.3234 | 0.1366 |
| External causes of morbidity and mortality | 0.7226 | 0.0979 | 0.2934 | 0.1053 |
| Factors influencing health status and contact with health services | 0.7591 | 0.0960 | 0.3077 | 0.1427 |
| (EXT) Injury, poisoning and certain other consequences of external causes | 0.7378 | 0.0657 | 0.2684 | 0.0930 |
| (BEH) Mental and behavioural disorders | 0.7656 | 0.0859 | 0.3175 | 0.1452 |
| (NEO) Neoplasms | 0.7571 | 0.0826 | 0.3028 | 0.1208 |
| (GYN) Pregnancy, childbirth and the puerperium | 0.9282 | 0.0856 | 0.6344 | 0.1879 |
| (LAB) Symptoms, signs and abnormal clinical and laboratory findings, not elsewhere classified | 0.7375 | 0.0779 | 0.2609 | 0.0903 |
| (RHEU) Systemic connective tissue disorders | 0.7334 | 0.0915 | 0.2730 | 0.1231 |

**Explaining output of POPDx**

We obtained feature relevance scores over 1,538 phenotypes using DeepLIFT (**Method**). For individual predictions, we visualize the features with a polar plot where the radial values represent their contribution scores. Because of the high dimensionality of patient features in our dataset, all the 38,663 variables are organized into 20 data subgroups that summarize different categories of UK Biobank data (Table S1). For the phenotype atherosclerosis (440.0 Diseases of arteries, arterioles, and capillaries), Figure 6A shows a polar chart for three true-positive patients, who share a similar pattern of feature importance. In particular, features from medical history, education and employment status, and physical measures are important. Figure 6B shows 5 categories of diseases for which we computed the max value of the importance scores of the selected data subgroups. Features in the lifestyle subgroup are critical for the recognition of endocrine, nutritional, and metabolic diseases. Patients' medical history is important for the recognition of phenotypes originating in the perinatal period. For the external causes of morbidity and mortality, the features related to mental health status are essential.

**DISCUSSION**

As large databases of patient records become available for research, associating precise phenotypes with patients becomes critical. Our method comprehensively and simultaneously scores hundreds to thousands of phenotypes. Automated phenotyping of a patient, even for a single disease, faces two central challenges: variations in the syntax and semantics of health records (different electronic systems, lack of standards for interoperability among hospitals, etc.), and patient-to-patient variability in the clinical manifestations of the diseases [32,33]. Most existing computational approaches to phenotype recognition are built by hand and model a small number of clinical, pathological, and laboratory attributes of patients [34,35]. These methods do not easily generalize to cover the whole disease ontology [36,37]. In addition, no methods using machine learning have been able to recognize phenotypes for which there are new or no training examples in the UK Biobank. Our integrated analysis allows us to make some progress in recognizing such examples.

The UK Biobank cohort is a long-tailed dataset with heavy class imbalance (Figure 2A). POPDx provides robust and scalable recognition of phenotypes; it performs quite well on common phenotypes for which there are many examples, and gracefully degrades its performance down to phenotypes with zero examples. The ROC curves (Figure S6) display the trade-off between sensitivity and specificity and can be helpful when considering the cut-off threshold to identify the diseased cohorts. The AUPRC (Figure 4) is prevalence dependent and less optimistic when the phenotype prevalence is low. Our framework significantly improved the AUPRC for unseen and rare phenotypes (<10 cases in training) by 218% and 151% compared to the logistic regression model. If a clinical team wants to identify patients for a phenotype with very low prevalence, our model on average doubles the ability to find the positive cases in the UK Biobank. When high specificity is desired over sensitivity in a clinical setting, more expert filtering can be used to detect the false positives. Our model provides a better starting point for the clinicians. This is encouraging since POPDx makes rare disease imputation feasible. Our method takes advantage of the non-linear correlation structure of the patient features to assign multiple diagnostic labels. In contrast, conventional machine learning methods such as random forest are unable to recognize phenotypes that are not present in the training dataset, inevitably limiting their applications in recognition of new phenotypes. Table 2 shows that the mean and standard deviation (SD) of AUROC and AUPRC are similar for certain disease categories (e.g., PERIN, CONGEN). This might imply that some of the phenotypes in those disease categories are close in the embedding space and exist on overlapping patients. In the future, we can explore other aspects of phenotyping such as disease comorbidity to explain specific patterns we have observed in this study.

Our framework incorporates structural knowledge of the disease ontology by embedding disease relationships in a low-dimensional space. The recognition of unseen and rare phenotypes is enabled by explicitly providing information about the network relationships of these phenotypes to others which are well-represented in our data set. This phenotype ontology embedding preserves proximity relations, so two representations of nearby phenotypes are embedded in similar locations. In addition to the disease ontology, our method leverages semantic information about phenotypes by including textual information (embedded by BioBERT [24]) that provides further context for the phenotype and its distance from other phenotypes. In addition to the widely used BioBERT, other BERT models trained on slightly different biomedical data such as PubMedBERT [38], BioClinicalBERT [39], and ClinicalBERT [39] can also serve our objective well. The POPDx makes it simple to train with different types of phenotype embeddings. Either the NLP-based or ontology-based embedding provides meaningful correlations of the phenotypes which can be demonstrated by their competitive performance in recognizing uncommon conditions. The combination of structure-based representations from the disease ontology with the contextual embeddings of phenotype text descriptions provides complementary information. The t-SNE representations (Figure 3A) of the joint embeddings demonstrate that our method preserves the separations of major disease types.

In our study, we assume that the collected ICD-10 codes and the associated phecode membership can be reliable representations of the underlying health state of the individuals in the UK Biobank resource. The faithfulness of diagnosis code can be compromised by several sources of errors. The complexities of the ICD coding system and a short time available for clinicians to match the patients with all the ICD-10 codes may cause inappropriate coding and variations in judgments [40,41]. Preferably, POPDx could be validated against a true "gold standard" manually contributed by the physicians. However, this is not practicable given the constraints inherent in the data source.

In clinical research, phenotype labels such as ICD-10 codes and phecodes enable an initial selection of patient cohorts [42]. We anticipate that POPDx will allow researchers to assemble patient cohorts beyond ICD-10 based search strategy, addressing the challenges of rare diseases and incomprehensive recognition of common phenotypes. With reliable identification of patients with phenotypes, we can use the genotype information present in the UK Biobank to seek genetic associations. We can also use known genetic associations between phenotypes to add an additional element to our embedding to help with phenotype recognition (i.e., a third element to our embedding in addition to the text and ontology structure). Our results demonstrate that our model's DeepLIFT feature relevance scores [29] can offer some insights to explain the assignment of 1,538 phenotypes. Our results with DeepLIFT show promise but they may not provide sufficiently clear justification for the reasons a feature contributes to a phenotype. For example, chest pain felt outside physical activity has a high importance score for the phecode 335.0 (hereditary/degenerative nervous conditions). While this may be reasonable, the mechanistic connection between these concepts is not clear.

The algorithm is implemented in Python (https://github.com/luyang-ai4med/POPDx).

**CONCLUSIONS**

The POPDx framework was developed for multi-phenotype recognition with heterogeneous patient data in the UK Biobank. Our model outperforms other existing methods on recognizing a comprehensive set of non-existent, rare, and common phenotypes in training. While we demonstrate our framework on UK Biobank, the model can be applied to any biobank-related records.

## ACKNOWLEDGEMENTS

We would like to thank Stanford sherlock cluster for high-performance computing (HPC) and access to the long-term data storage through the OAK research filesystem and compute nodes (GPUs, CPUs).

## FUNDING

This work was supported by the National Institutes of Health grant number GM102365 and Chan-Zuckerberg initiative. LY is supported by Agilent and the Chan-Zuckerberg Biohub.

## COMPETING INTERESTS

The authors have no competing interests to declare.

## CONTRIBUTION

LY extracted data, carried out the experiments, and conducted the analyses. LY, SW, and RBA conceived the original idea. RBA supervised the project. All authors revised and approved the final manuscript.

## DATA AVAILABILITY

The data used are anonymized and available from the UK Biobank through approved access (https://www.ukbiobank.ac.uk/enable-your-research/apply-for-access).

**FIGURES**

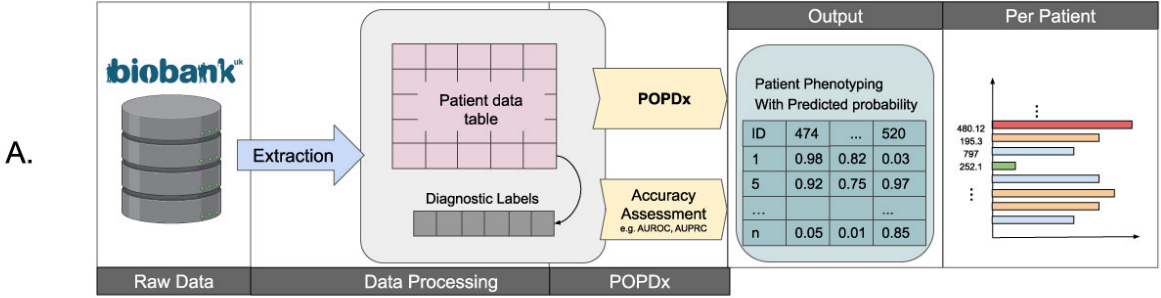

A.

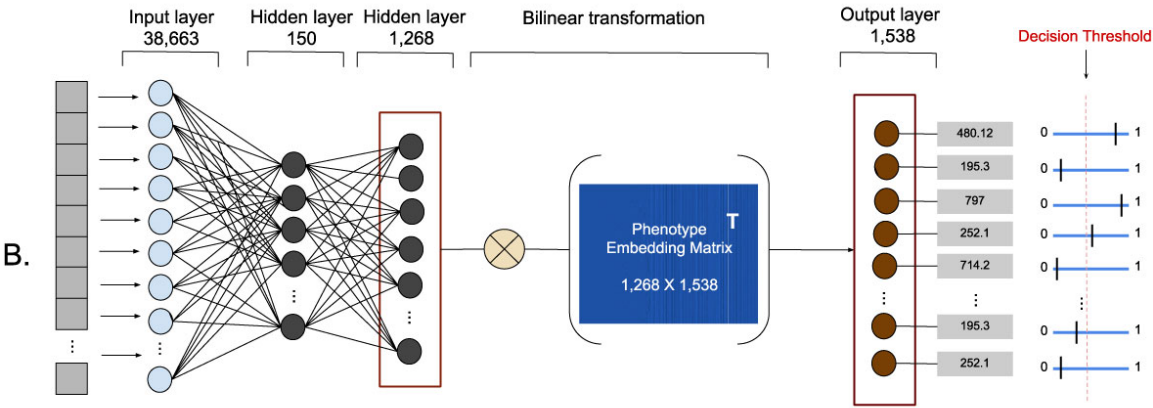

B.

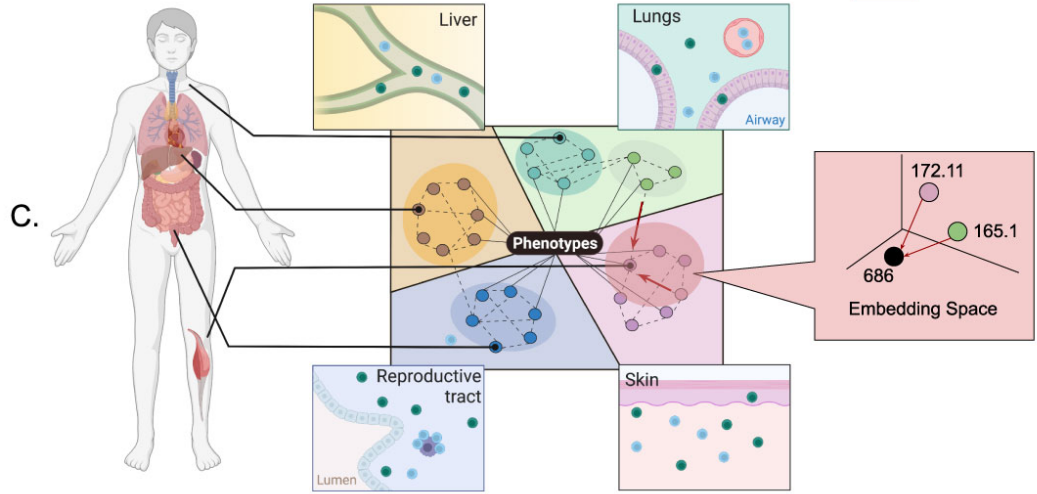

C.

Figure 1. POPDx overview. (A) The flowchart of patient phenotyping in UK Biobank includes raw data extraction, data processing, POPDx development, and result evaluation. After extracted and preprocessed the raw data from UK Biobank, we obtained vectors of features for all patients as shown in the patient data table and a vector of diagnostic labels of phecodes for each patient. We developed POPDX to encode the patient features that are eventually used to perform phenotype recognition. The output of POPDx is a matrix of probabilities which represents the likelihoods of all phecodes for all the patients. We evaluated the accuracy of POPDx by AUROC and AUPRC. (B) The architecture of POPDx is a bilinear model that leverages an embedding matrix of 1,538 phecodes. A total number of 38,663 features per patient were input into POPDx. The structure of POPDx includes an input layer, two hidden layers each with 150 and 1,268 neurons, and a bilinear transformation through an embedding matrix of the phenotypes. The output layer is the probability distribution that this patient is assigned with each phecode label. The predicted labels are either 0 or 1 based on the decision threshold, illustrated as the vertical dashed line.  (C) POPDx embeds the phenotypes into low-dimensional space. The hierarchical tree-structured representation of the phenotypes is utilized. For simplicity, we only show 4 categories of diseases. Even if the phecode does not have any patient examples in the training data (686: local infections of skin and subcutaneous tissue), POPDx can leverage its relation to other phecodes (172.11: melanomas of skin, 165.1: cancer of bronchus and lung) in the embedding space. Figure adapted from "Distribution of TRM Cells", by BioRender.com (2022). Retrieved from https://app.biorender.com/biorender-templates.

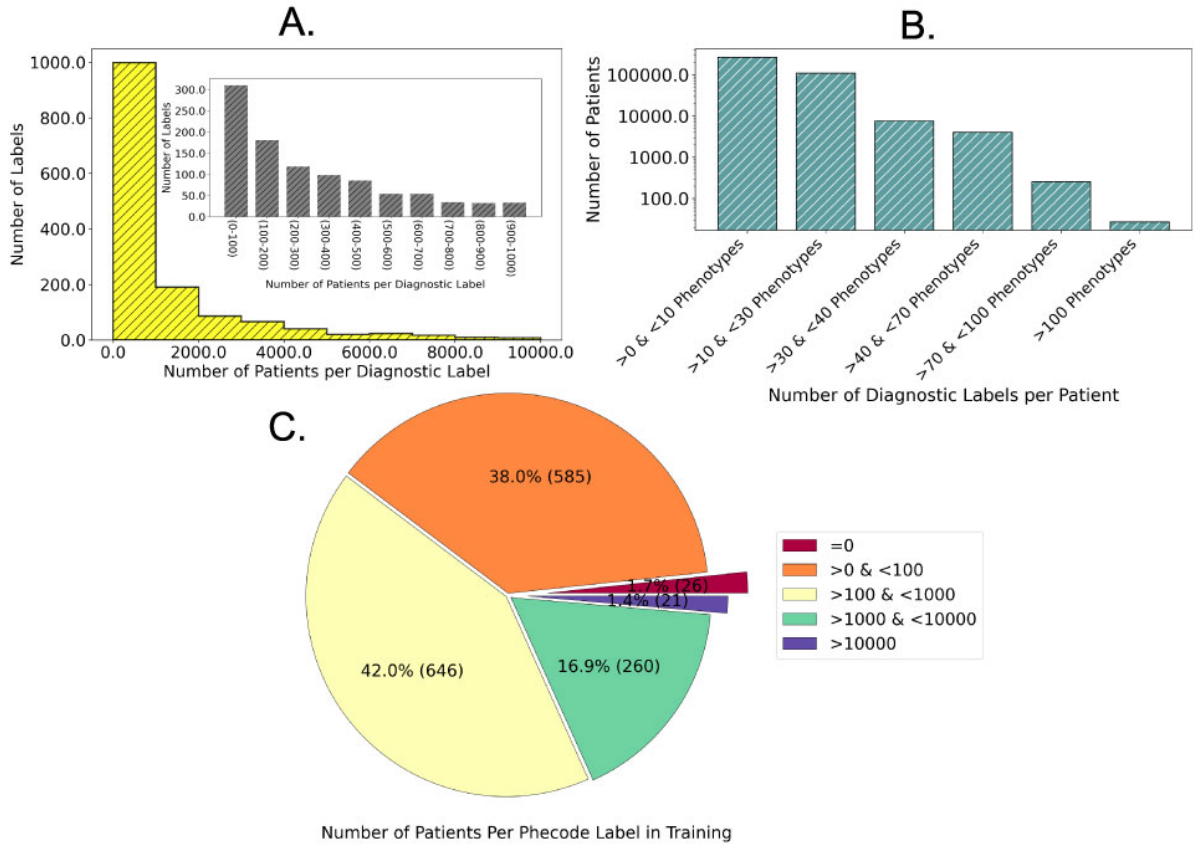

Figure 2. Diagnostic label statistics. (A) The UK Biobank has a long-tailed distribution of phecode labels. The x-axis is the number of patients per phecode label and y-axis is the number of labels. Most of the phecode labels have fewer than 1000 patients in the UK Biobank. (B) The patients in UK Biobank are associated with multiple phenotype labels. The x-axis is the groups of phecode counts and log-scale y-axis is the number of patients. The majority of the patients have fewer than 30 phecodes labels. (C) The phenotypes are categorized based on the number of training samples. The exploded pie chart shows the relative abundance of phecodes based on the number of patients in training.

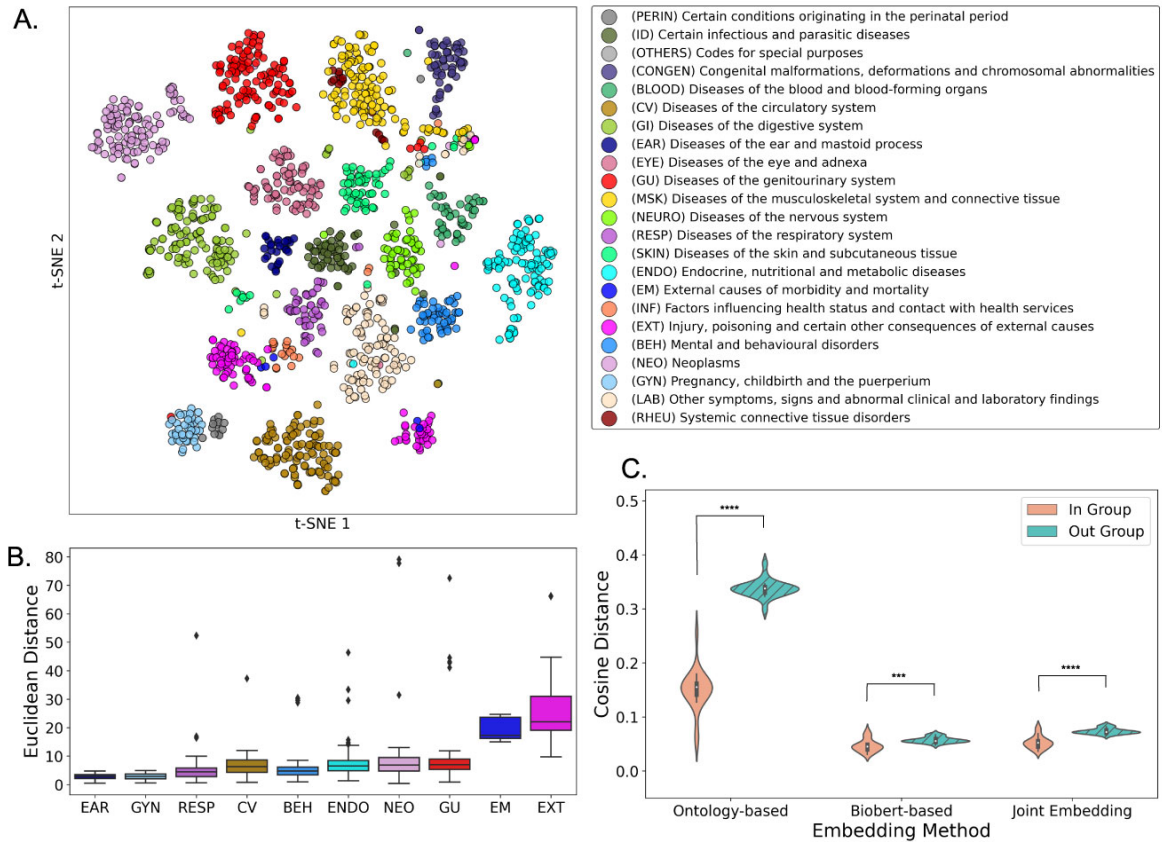

Figure 3. Phenotype embeddings. (A) The *t*-SNE plot shows the segregation of phenotypes into different disease categories using the joint structure and semantic embedding method. The legend associates a color to each phenotype category based on the hierarchical tree-structure of ICD-10. (B) The Euclidean distance of phecodes to the cluster center of each disease category in the t-SNE plot. (C) The similarity analysis of phenotype groups embedded with three different methods is presented as cosine differences within groups and between groups.

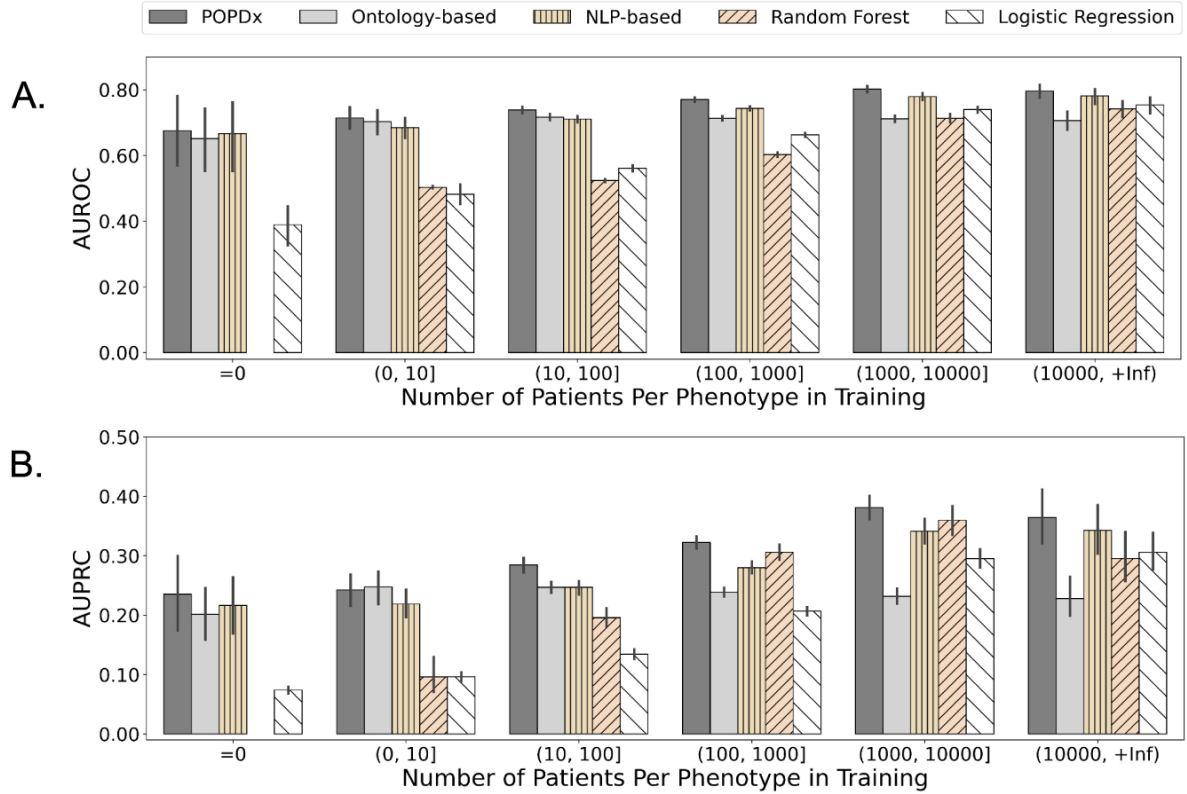

Figure 4. POPDx can recognize phenotypes that are not present in the training set. Bar plots comparing POPDx and other methods in terms of (A) AUROC and (B) AUPRC on the test set. POPDx presents competitive performance across all the groups of phenotypes.

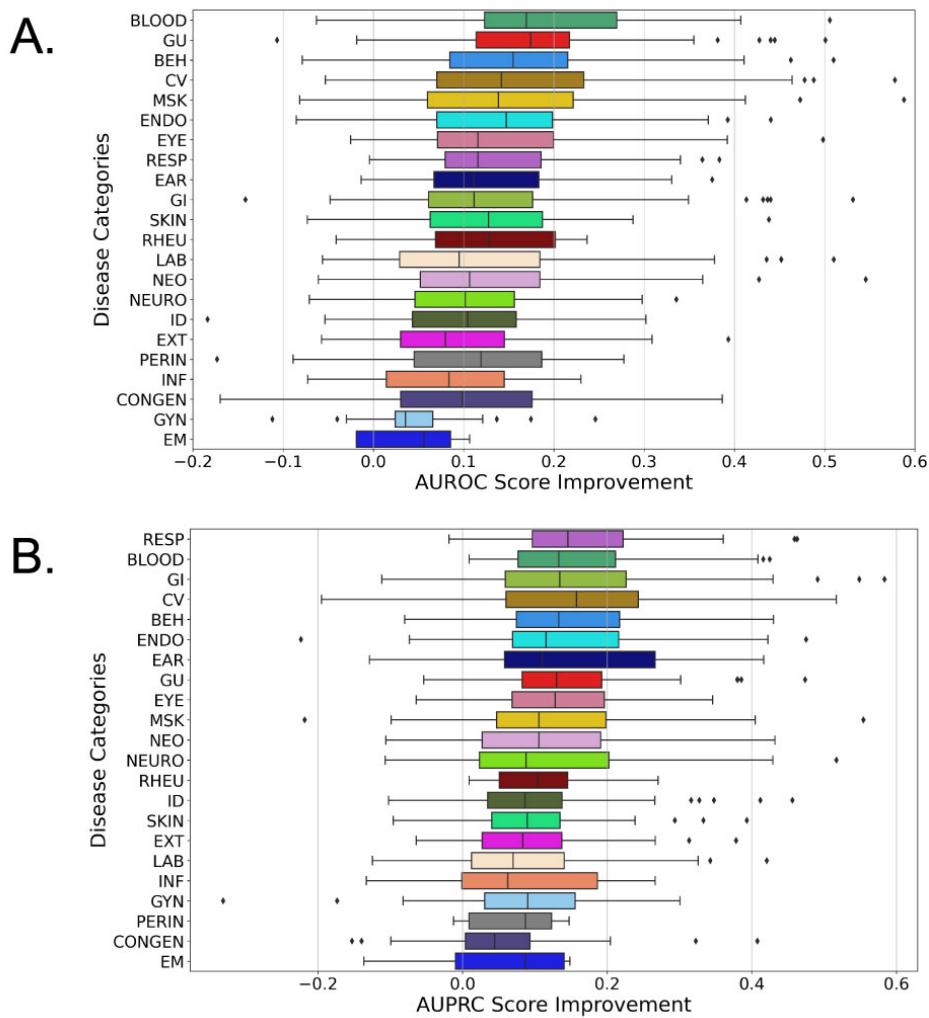

Figure 5. POPDx improves disease recognition compared with logistic regression across 22 disease categories. (A) The AUROC scores for all the disease categories are substantially improved compared to the logistic regression (LR) baseline. The x-axis is the improvement of AUROC score by POPDx. The y-axis represents different disease categories. (B) The AUPRC scores for all the disease categories are substantially improved compared to the logistic regression (LR) baseline. The x-axis is the improvement of AUPRC score by POPDx. The y-axis represents different disease categories.

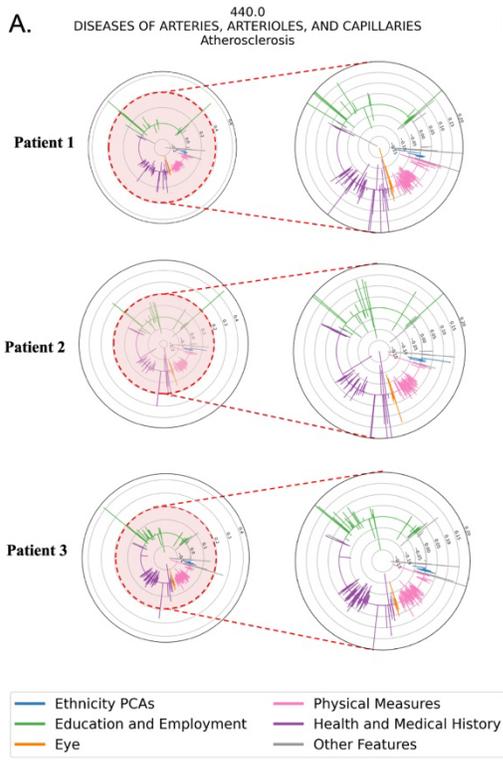

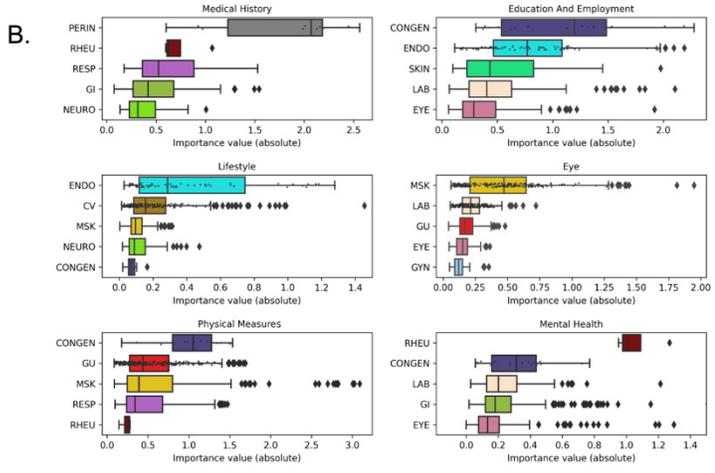

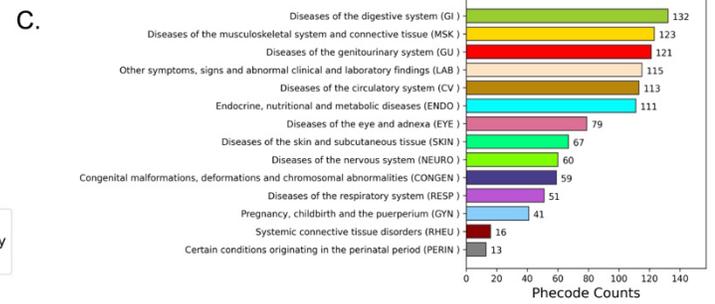

Figure 6. POPDx explainability across different disease categories. (A) Feature relevance profile for phenotype 440.0. The polar plots summarize feature importance for three true-positive patients with atherosclerosis. The polar plots on the right represent the zoomed-in region as highlighted in red to aid the eye. The legend associates a line color to a feature subgroup. Radial value in the polar plots is feature importance magnitude from DeepLIFT. All features are analyzed into different subgroups to show the consistency of the POPDx interpretability. (B) Additive contributions of individual features (medical history, education and employment, lifestyle, eye, physical measures, mental health shown here) to the outputs of POPDx. Each color is associated with a disease category which is consistent with the colors of disease categories in panel C. X-axis specifies the importance value of the corresponding feature subgroups (higher importance to the right). Y-axis is 5 disease categories sampled based on the median importance values from DeepLIFT. (C) A horizontal bar plot of the number of phecodes in the disease categories. The colors in (B) and (C) are consistently used to represent different categories of phenotypes.

**Supplemental**

Table S1. Feature categories

| Data Subgroup | Data Category in UK Biobank Showcase | The Number of Variables as POPDx Input | |
|---|---|---|---|
| Medical history | Summary Operations | 5527 | 13795 |
| | Medical conditions | 2377 | |
| | Death register | 2008 | |
| | Cancer register | 1528 | |
| | Operations | 1216 | |
| | Summary Administration | 937 | |
| | Claudication and peripheral artery disease | 43 | |
| | Pain | 42 | |
| | Medical information | 36 | |
| | General health | 20 | |
| | Chest pain | 16 | |
| | Stroke outcomes | 12 | |
| | Cancer screening | 8 | |
| | Sleep | 8 | |
| | Myocardial infarction outcomes | 6 | |
| | Sexual factors | 6 | |
| | Ongoing characteristics | 5 | |
| Education and Employment | Employment | 10105 | 11739 |
| | Employment history | 1619 | |

|  | Education | 15 |  |
|---|---|---|---|
| Physical measures | Autorefraction | 3397 | 5807 |
| | Acceleration intensity distribution | 550 | |
| | Acceleration averages | 409 | |
| | Body composition by impedance | 352 | |
| | Accelerometer calibration | 177 | |
| | Body size measures | 161 | |
| | Carotid ultrasound | 144 | |
| | Physical activity | 138 | |
| | Spirometry | 136 | |
| | Accelerometer wear time duration | 89 | |
| | Raw accelerometer statistics | 88 | |
| | ECG during exercise | 57 | |
| | MET Scores | 38 | |
| | Arterial stiffness | 34 | |
| | Hand grip strength | 26 | |
| | Breathing | 8 | |
| | ECG at rest, 12-lead | 3 | |
| Medication | Medications | 3709 | 3721 |
| | Cannabis use | 12 | |
| Mental health | Mental health | 192 | 605 |
| | Depression | 125 | |
| | Anxiety | 112 | |
| | Self-harm behaviors | 40 | |
| | Unusual and psychotic experiences | 37 | |
| | Mania | 26 | |
| | Mental distress | 26 | |
| | Summary Psychiatric | 24 | |
| | Happiness and subjective well-being | 23 | |
| Ethnicity PCAs | Genotyping process and sample QC | 588 | 588 |
| | Diet | 155 | |
| | Residential air pollution | 134 | |
| | Household | 67 | |
| | Electronic device use | 59 | |
| | Residential noise pollution | 55 | |
| Lifestyle | Sun exposure | 34 | 575 |
| | Social support | 32 | |
| | Sleep | 22 | |
| | Other sociodemographic factors | 12 | |
| | Diet by 24-hour recall | 3 | |
| | Ongoing characteristics | 2 | |
| | Blood count | 313 | |
| | Blood sample collection | 22 | |
| Biological sample | Saliva sample collection | 8 | 355 |
| | Urine sample collection | 8 | |
| | Urine processing | 4 | |
| Eye exams | Intraocular pressure | 115 | 269 |
| | Eyesight | 64 | |
| | Visual acuity | 55 | |
| | Eye surgery/complications | 35 | |
| Early life factors | Early life factors | 244 | 244 |
| Addiction | Smoking | 117 | 165 |
| | Addictions | 43 | |
| | Medical information | 5 | |
| Cognitive assessment | Fluid intelligence / reasoning | 92 | 142 |
| | Reaction time | 20 | |
| | Prospective memory | 17 | |
| | Numeric memory | 6 | |
| | Procedural metrics | 4 | |
| | Word production | 3 | |
| Women's health | Summary Maternity | 88 | 131 |
| | Female-specific factors | 43 | |
| Family history | Family history | 124 | 124 |
| Traumatic history | Traumatic events | 114 | 114 |
| Baseline characteristics | Reception | 66 | 113 |
| | Baseline characteristics | 25 | |
| | Ethnicity | 22 | |
| Alcohol use | Alcohol use | 56 | 103 |
| | Alcohol | 47 | |

| Hearing exams | Hearing | 43 | 43 |
|---|---|---|---|
| Men's health | Male-specific factors | 16 | 16 |
| Dental health | Mouth | 14 | 14 |

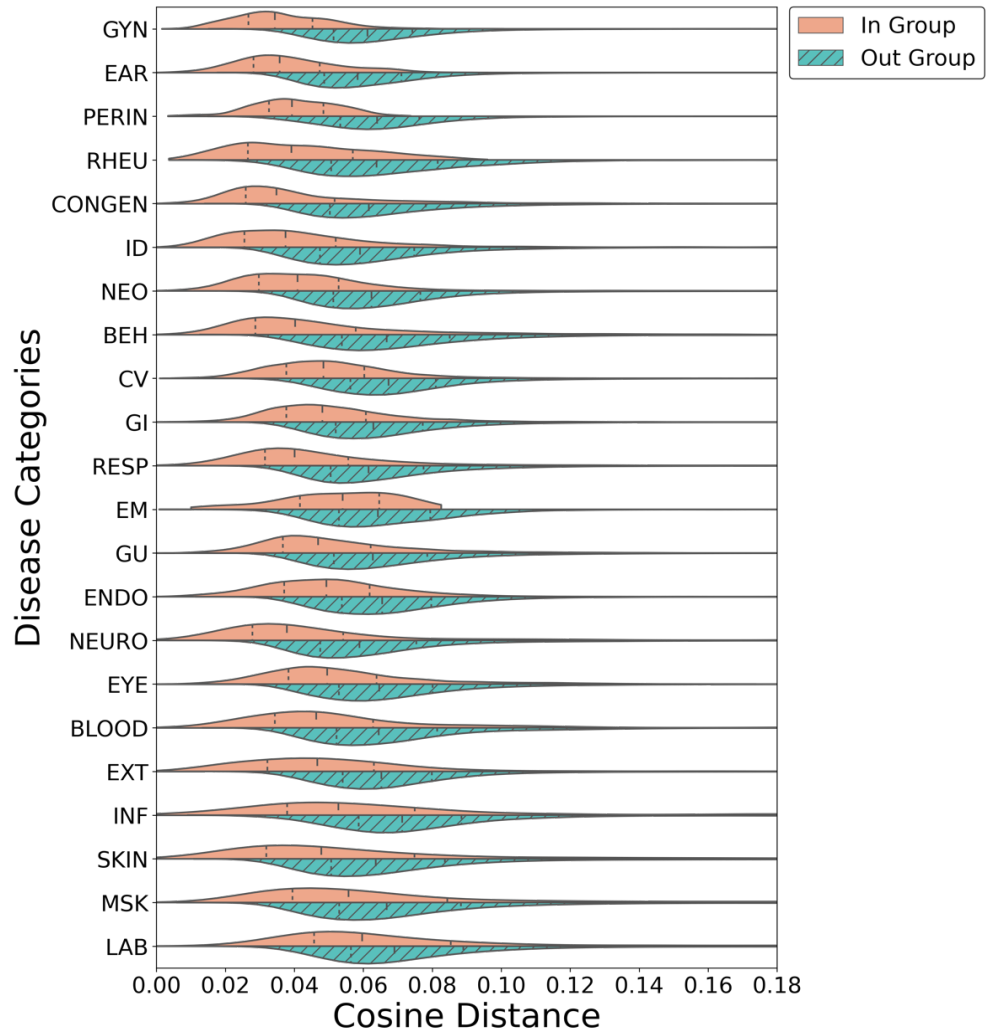

Figure S2. Phenotypes dissimilarity analysis across different disease categories. The cosine distances of phenotypes in different disease categories are represented as in-group (intra-) and out-group (inter-) distances.

## ID

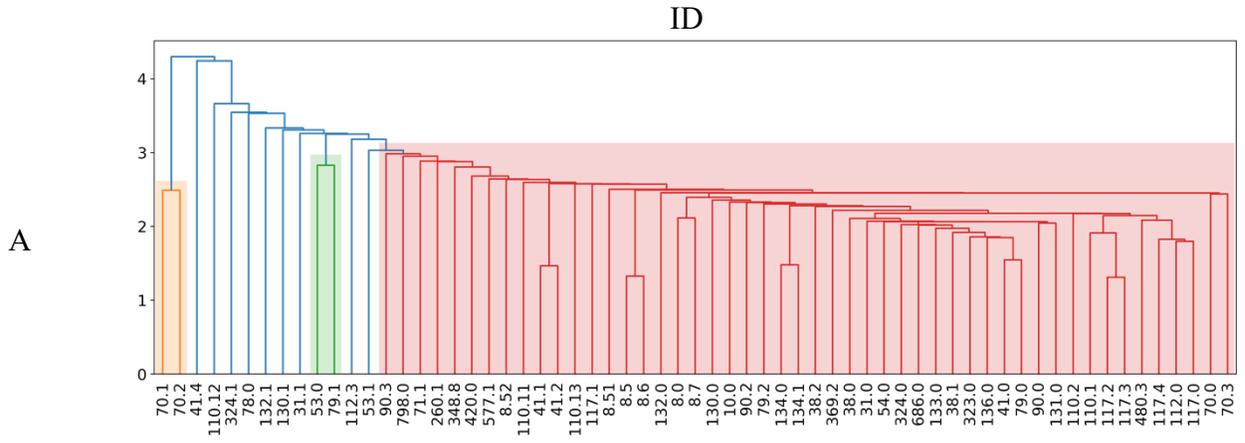

## BLOOD

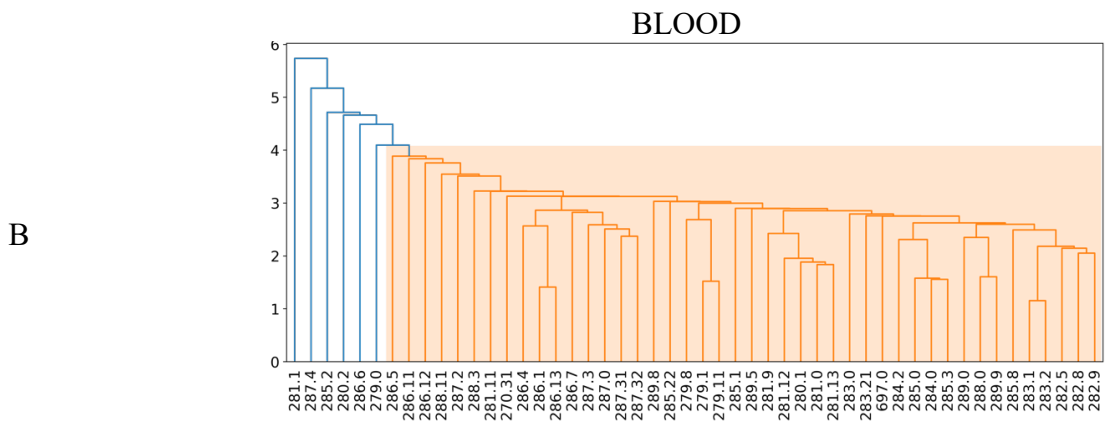

## BEH

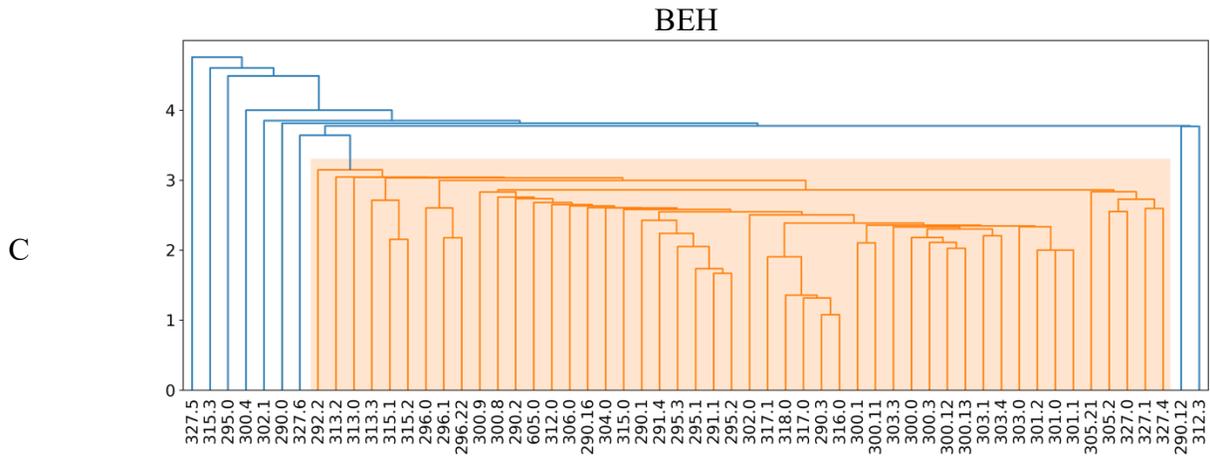

D

## RHEU

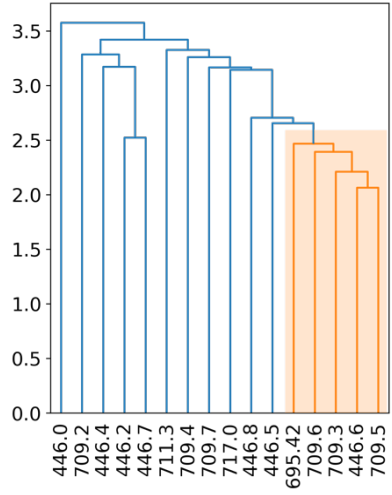

## RESP

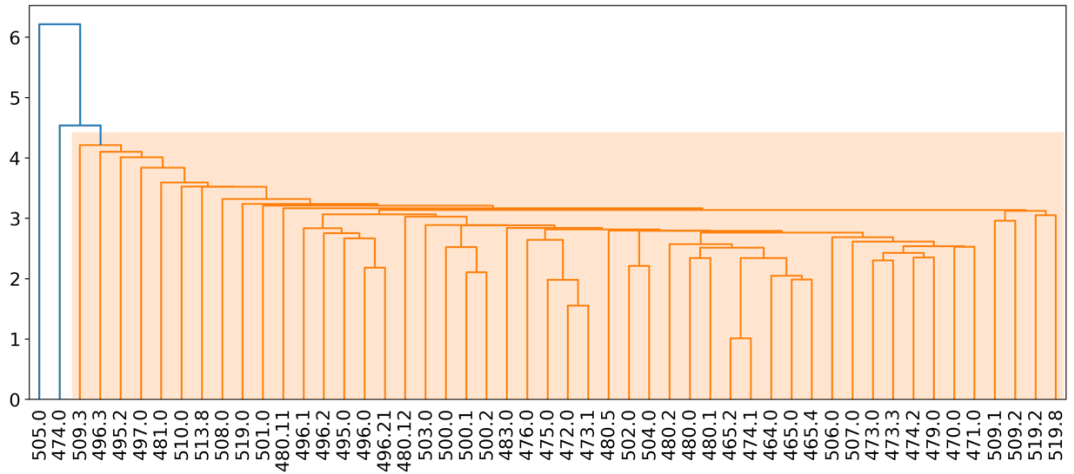

E

## NEURO

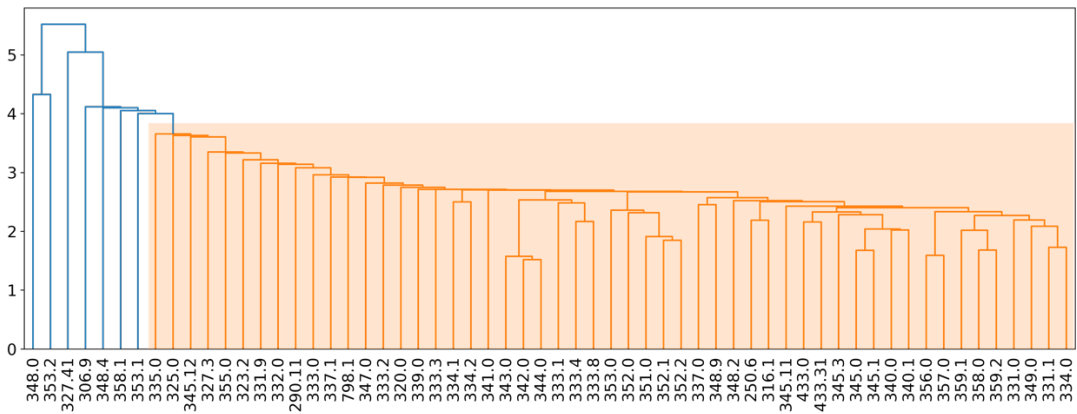

F

G                                    INF

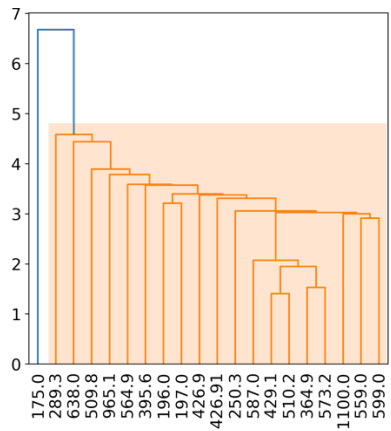

GU

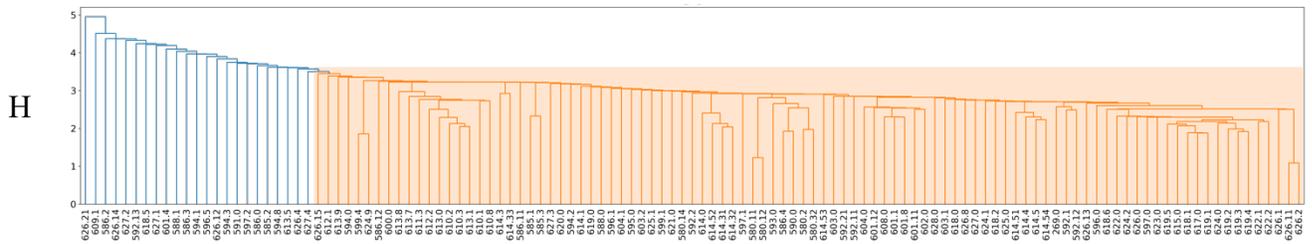

H

MSK

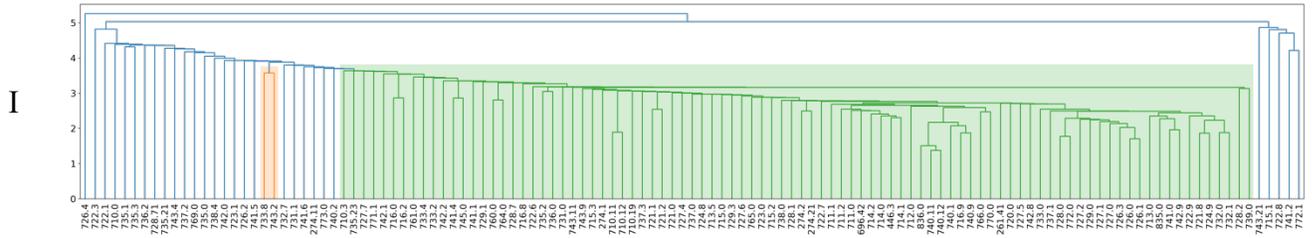

I

GYN

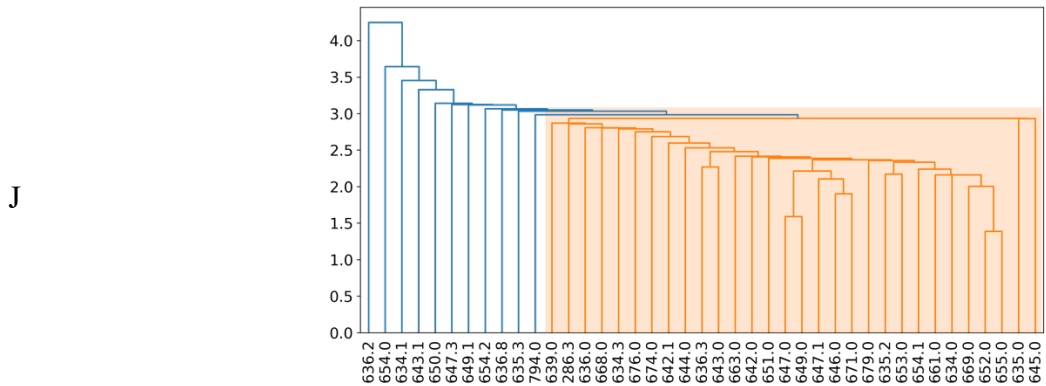

J

SKIN

K

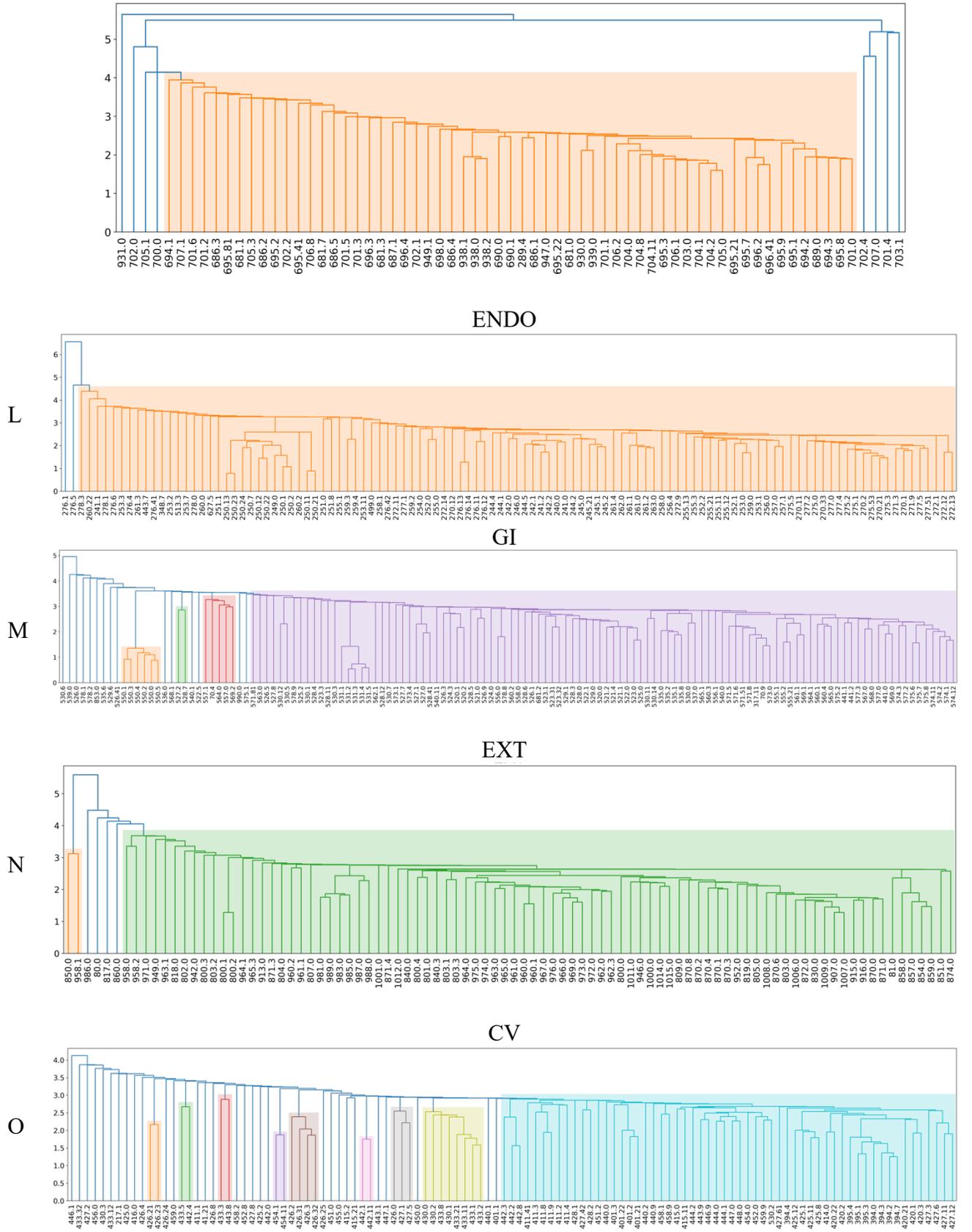

ENDO

GI

EXT

CV

P

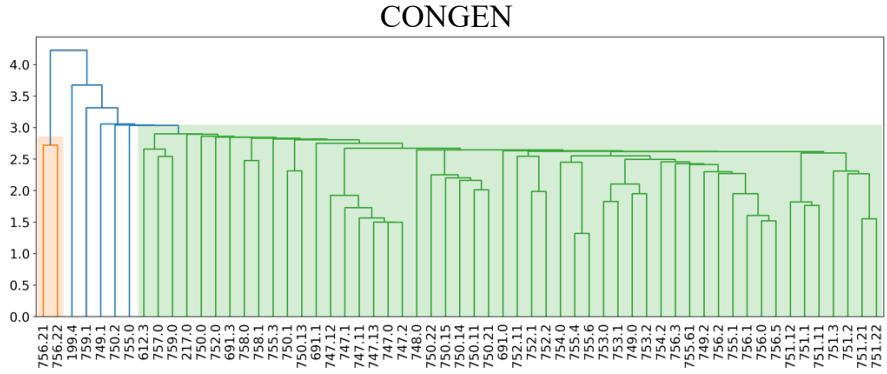

Q

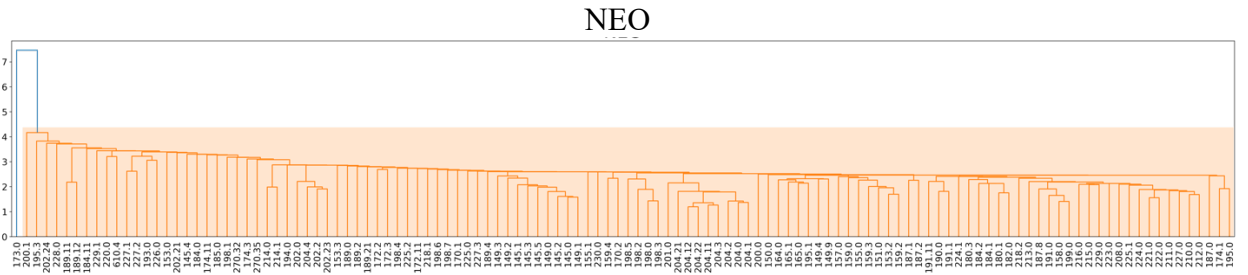

R

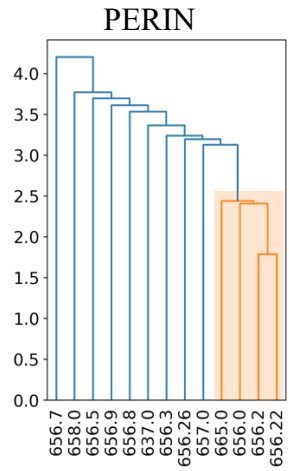

S

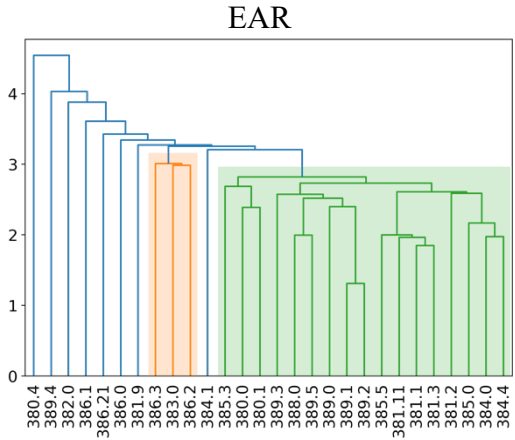

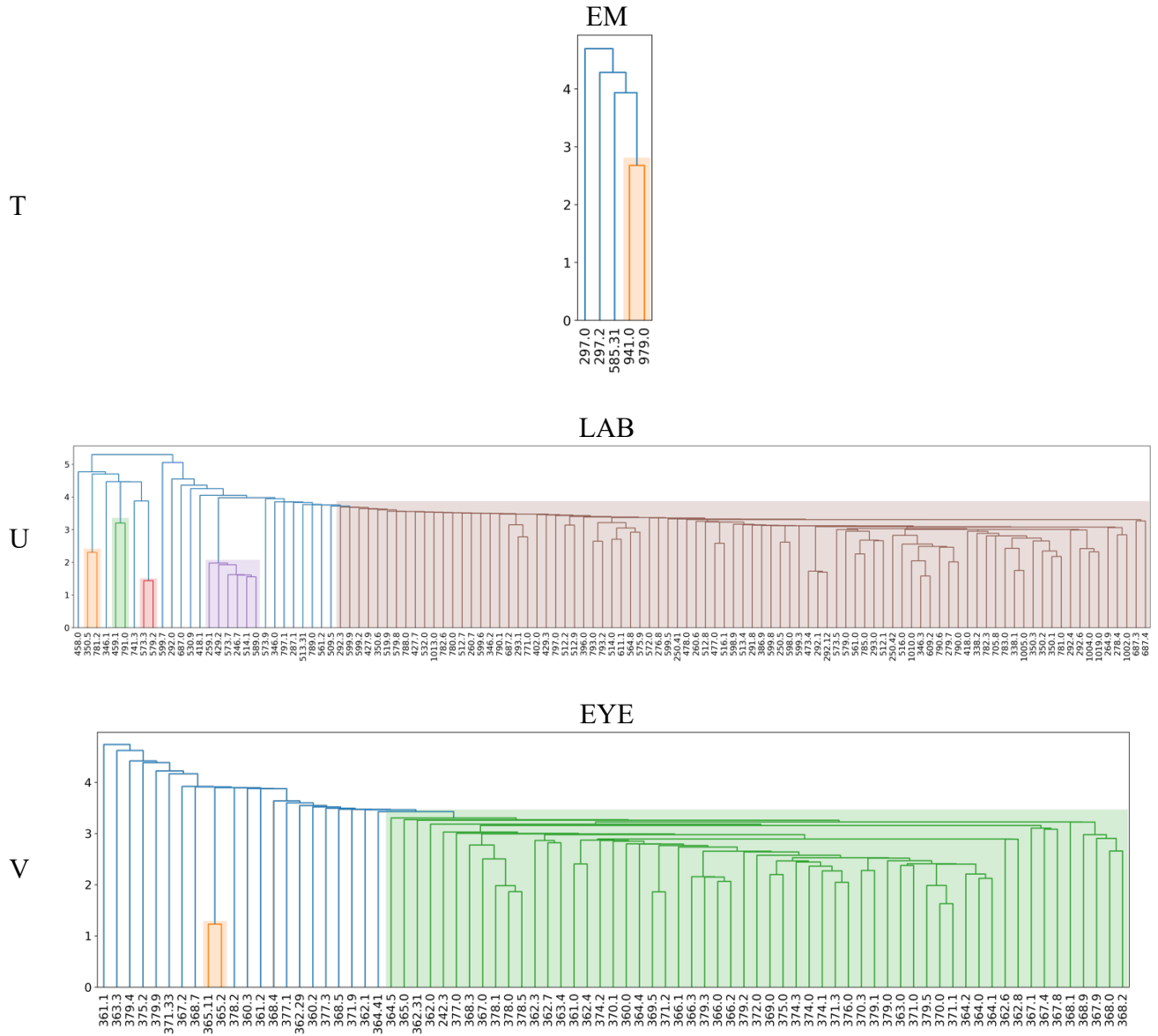

Figure S3. Dendrograms of phenotype representations in 22 disease categories. The Phecodes are clustered based on the embedding similarity from POPDx.

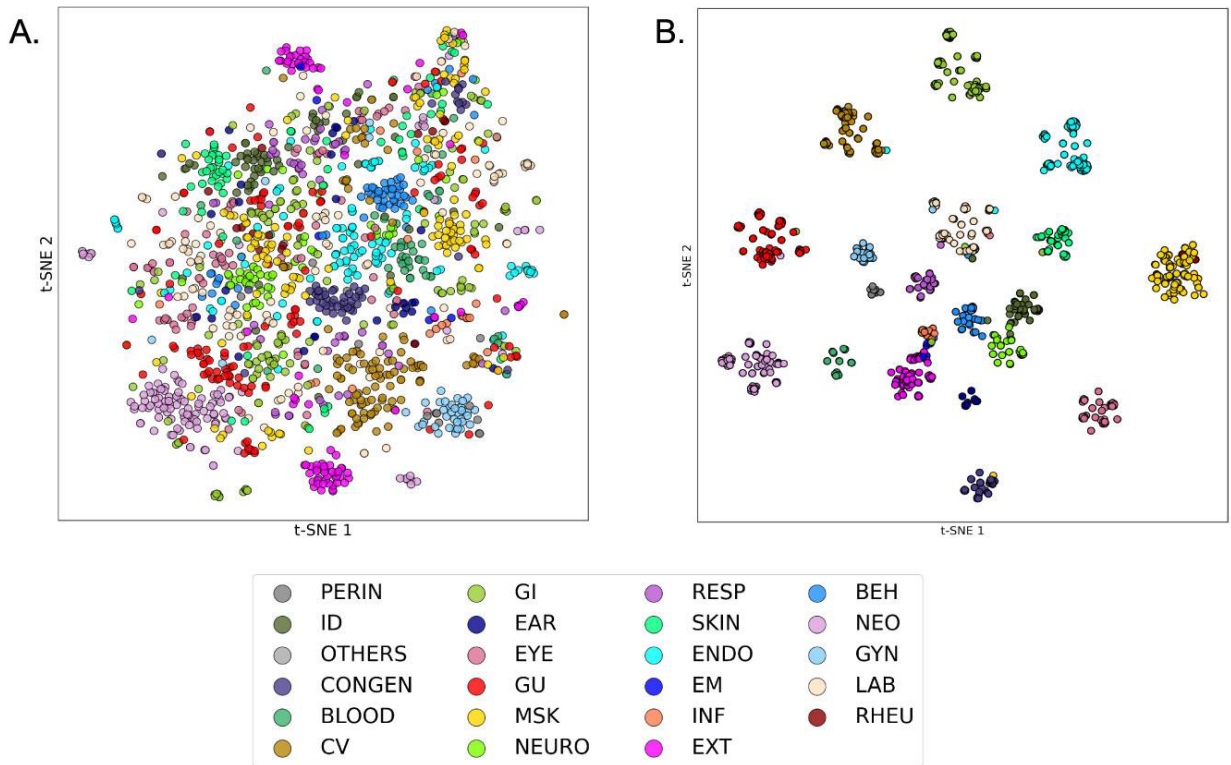

Figure S4. The structure-based embeddings and NLP-based embeddings of phenotypes. (A) The *t*-SNE plot shows less segregation of phenotypes using the semantic embedding method. (B) The *t*-SNE plot shows the segregation of phenotypes into different disease categories using the structure-based embedding method.

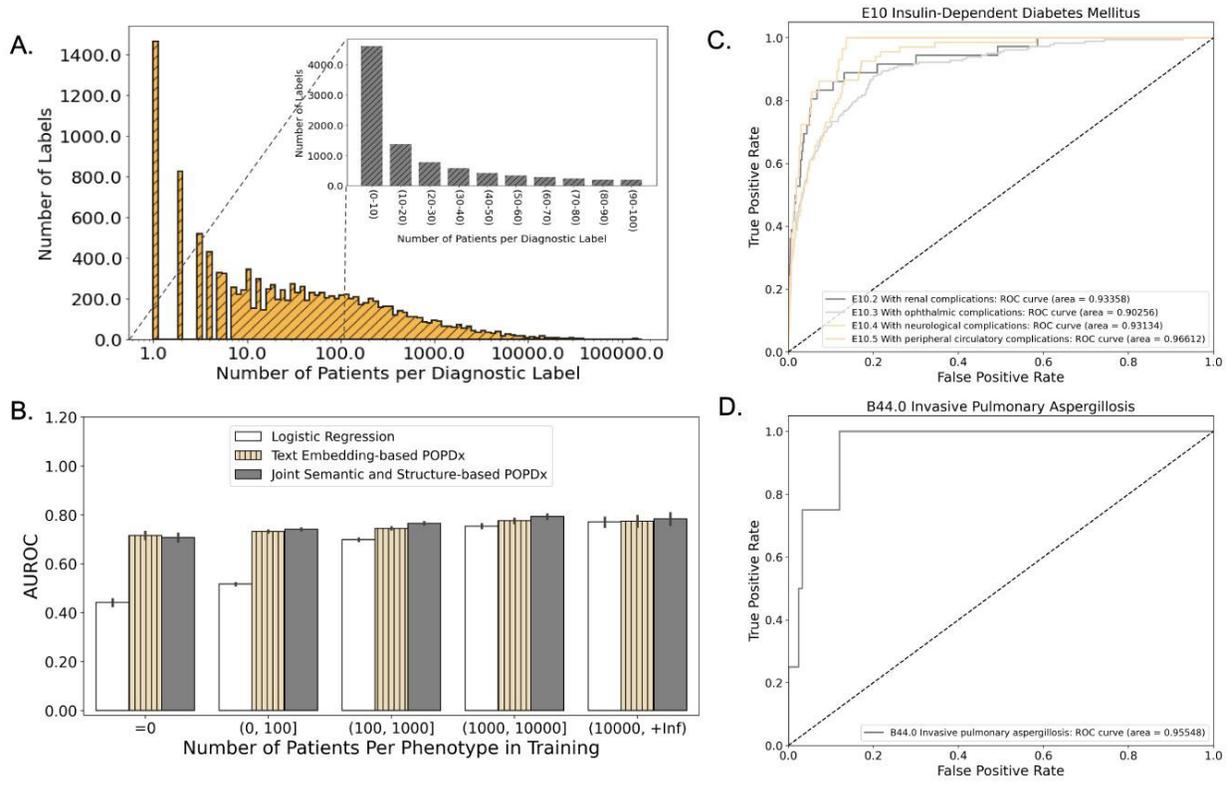

Figure S5. The POPDx performance for patient phenotyping of 12,803 ICD-10 codes in the UK Biobank study. (A) The long-tailed distributions of the ICD-10 diagnostic labels present a challenging dataset with heavy data imbalance. (B) A bar plot comparing POPDx with other models in terms of AUROC scores. (C) The ROC curves for phenotypes related to insulin-dependent diabetes mellitus (seen phenotypes in training). (D) The ROC curve of a rare phenotype in training.

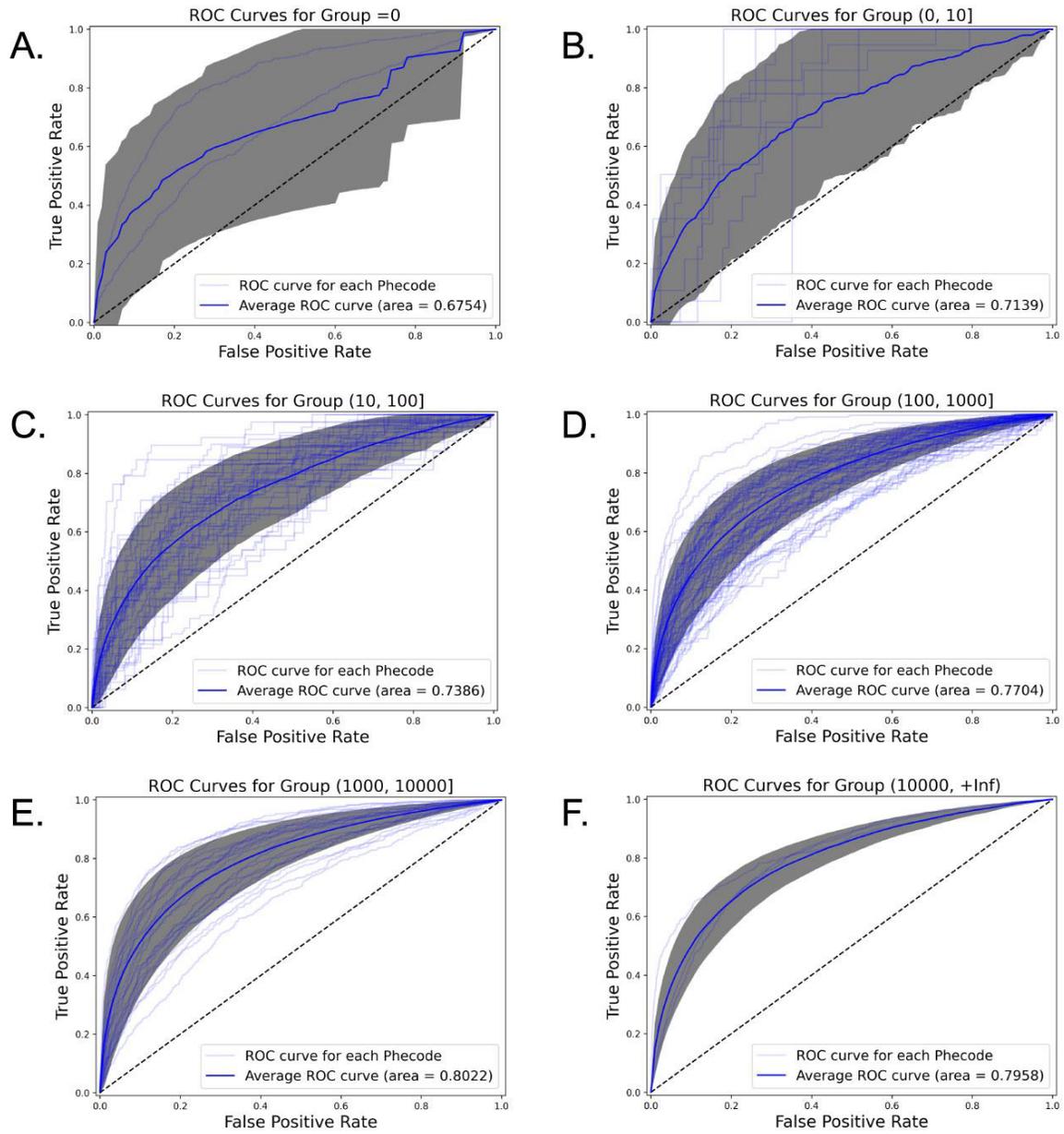

Figure S6. The POPDx performance on test sets for patient phenotyping of 1,538 Phecodes in the UK Biobank study. ROC curves across six groupings of Phecodes are represented respectively: (A) 0 samples (B) (0, 10] samples (C) (10, 100] samples (D) (100, 1000] samples (E) (1000, 10000] samples (F) (10000, +Inf) samples in the training set. A random selection of the ROC curves is shown to aid the eye.